\documentclass[submission,copyright,creativecommons]{eptcs}
\providecommand{\event}{QAPL 2017} 
\usepackage{breakurl}             
\usepackage{underscore}           

\usepackage{amsmath,amssymb}
\usepackage{arrows}
\usepackage{graphicx}
\usepackage{mathrsfs}
\usepackage{color}

\newtheorem{proposition}{Proposition}

\newtheorem{example}{Example}
\newtheorem{definition}{Definition}

\newcommand{\AND}{\mbox{$\;\;\wedge\;\;$}}
\newcommand{\false}{\mbox{ff}}

\newcommand{\true}{\mbox{tt}}

\def\FlyFast#1{{\sf FlyFast#1}}
\def\Abf#1{{\sf PiFF#1}}
\newcommand{\SET}[1]{\{#1\}}
\newcommand{\aN}{^{(N)}}

\def\calA{{\cal A}}
\def\calI{{\cal I}}
\def\calL{{\cal L}}

\def\calO{{\cal O}}
\def\calP{{\cal P}}
\def\calS{{\cal S}}
\def\calU{{\cal U}}
\def\calV{{\cal V}}
\newcommand{\tuple}[1]{{\langle #1 \rangle}}

\def\act#1{{\calA}_{#1}} 	
\def\dfas{:=}			

\def\frc{\textsf{frc}\,}
			
\def\otm{\mathbf{K}}				
\def\os{\Delta}			
\def\semint#1#2{#1[\![#2]\!]}

\def\sep{|}				
\def\sc#1{\calS_{#1}}		

\def\attribute{\mathbf{attribute}}
\def\case{\mathbf{case}}
\def\float{\mathbf{float}}
	
\def\const{\mathbf{const}}	
\def\enum{\mathbf{enum}}	
\def\endfunc{\mathbf{endfunc}}	
\def\endupdate{\mathbf{endupdate}}	
\def\func{\mathbf{func}}		
\def\my{\mathbf{my}}	
\def\of{\mathbf{of}}

\def\attype{\mathbf{attype}}	
\def\update{\mathbf{update}}
\def\otherwise{\mathbf{otherwise}}
\def\with{\mathbf{with}}

\def\us#1{\calU^#1}
\def\vct#1{{\mathbf #1}}
\def\loc{\mathsf{loc}}

\def\rest{\mathbf{rest}}
\def\cond{\mathsf{cond}}
\def\synsum{\mathsf{SUM}}

\def\synprod{\mathsf{PROD}}
\def\carma{\textsc{Carma}}

\def\reals{\mathbb{R}}
\def\deriv{\noindent\hspace*{.20in}\vspace{0.1in}}

\def\hint#1#2{\newline#1\hspace*{.25in}\{$#2$\}\vspace{0.1in}\newline\hspace*{.20in}\vspace{0.1in}}

\title{Design  and Optimisation of the \FlyFast{} Front-end for Attribute-based Coordination}
\author{Diego Latella \qquad \qquad Mieke Massink 
\institute{Consiglio Nazionale delle Ricerche}
\institute{Istituto di Scienza e Tecnologie dell'Informazione ``A. Faedo''}
\email{Diego.Latella@isti.cnr.it, Mieke.Massink@isti.cnr.it}
}
\def\titlerunning{Designing Attribute-based \FlyFast{}}
\def\authorrunning{D. Latella \& M. Massink}

\begin{document}
\maketitle

\begin{abstract}
Collective Adaptive Systems (CAS) consist of a large number of interacting objects. 
The design of such systems requires scalable analysis tools and methods, which 
have necessarily to rely on some form of approximation of the system's actual behaviour.
Promising techniques are those based on mean-field approximation. 
The \FlyFast{} model-checker 
uses an on-the-fly algorithm for bounded PCTL model-checking of selected 
individual(s) in the context of very large populations whose global
behaviour is approximated using deterministic limit  mean-field techniques. 
Recently, a front-end for \FlyFast{} has been proposed which provides a modelling language, \Abf{} in the sequel, for the {\em Predicate-based} Interaction for \FlyFast{.} 
In this paper we present details of \Abf{} design and an approach to state-space reduction
based on probabilistic bisimulation for inhomogeneous DTMCs.
\end{abstract}

\section{Introduction}
Collective Adaptive Systems (CAS) consist of a large number of entities with 
decentralised control and varying degrees of complex autonomous behaviour. They form the basis of many modern {\em smart city} critical infrastructures. Consequently, their design requires  support from 
formal methods and scalable automatic tools based on solid mathematical foundations. 
In~\cite{LLM15a,LLM14}, Latella et al. presented a scalable mean-field model-checking procedure for verifying bounded  Probabilistic Computation Tree Logic (PCTL,~\cite{HaJ94}) properties of an  individual\footnote{The technique can be applied also to a finite selection of individuals; in addition, systems with several distinct types of individuals can be dealt with; for the sake of simplicity, in the present paper we consider systems with many instances of a single individual only and we focus in the model-checking a single individual in such a context.} in the context of a system consisting of a large number of interacting objects.  The model-checking procedure is implemented in the tool  \FlyFast{\footnote{{\tt http://j-sam.sourceforge.net/}}}.
The procedure performs on-the-fly, mean-field, approximated model-checking based on the idea of fast simulation, as introduced in~\cite{BMM07}. More specifically, 
the behaviour of a generic agent with $S$ states in a system  with a {\em large} number $N$ of instances of the agent at given step (i.e. time) $t$ 
is approximated by $\otm(\boldsymbol{\mu}(t))$ where $\otm(\vct{m})$ is 
the $S \times S$ probability transition matrix of 
an (inhomogeneous) DTMC and $\boldsymbol{\mu}(t)$ is a vector of size $S$ approximating
the mean behaviour of (the rest of) the system at  $t$;  each element  of $\boldsymbol{\mu}(t)$
is associated with a distinct state of the agent, say $C$, and gives an approximation of the fraction  of instances of the agent that are in state $C$ in the global system, at step $t$.
Note that such an approximation is a {\em deterministic} one, i.e. 
$\boldsymbol{\mu}$ is a {\em function} of the step $t$ (the exact behaviour of the rest of the
system would instead be a {\em large} DTMC in turn); note furthermore, that the above transition matrix
does not depend on $N$~\cite{LLM15a,LLM14}.

Recently, modelling and programming languages have been proposed specifically for autonomic computing systems and CAS~\cite{De+15,Bo+15}. Typically, in such frameworks, a system is composed of a set of independent {\em components} where a component is a process equipped also with a set of {\em attributes} describing features of the component. The attributes of a component  can be {\em updated} during 
its execution so that the association between attribute {\em names} and attribute {\em values}
is maintained in the dynamic {\em store} of the component. Attributes can be used in {\em predicates} appearing in language constructs for component interaction. The latter is thus typically modelled using
{\em predicate-based} output/input {\em multicast}, originally proposed in~\cite{Lat83}, and 
playing a fundamental role in the interaction schemes of languages like SCEL~\cite{De+15} and \carma{}~\cite{Bo+15}.
In fact, predicate-based communication can be used by components to dynamically organise themselves 
into ensembles and as a means to dynamically select partners for interaction. Furthermore,
it provides a way for representing component features, like for instance component location in space,
which are  fundamental for systems distributed in space, such as CAS~\cite{LoH16}.

In~\cite{CLM16a} we proposed a front-end modelling language for \FlyFast{} that provides constructs for dealing with {\em components} and {\em predicate-based interaction};  in the sequel, the
language---which has been inspired by \carma{---} will be referred to  as \Abf{,} which stands for for 
Predicate-based Interaction for \FlyFast{.}
Components interact via predicate-based communication. Each component consists of a behaviour, modelled as a  DTMC-like agent, like in \FlyFast{,} and a set of attributes. The attribute
name-value correspondence is kept in the current store  of the component.  
Actions are {\em predicate based multi-cast} output and input primitives; predicates are defined over attributes. 
Associated to each action there is also an (atomic) probabilistic store-update. 
For instance, assume components have an attribute named $\loc$ which takes values in the set of points
of a space, thus recording the current location of the component. The following action models a multi-cast via channel $\alpha$ to all 
components in the same location as
the sender, making it change location randomly: 
$\alpha^*[\loc=\my.\loc]\tuple{}\mathtt{Jump}$. Here $\mathtt{Jump}$ is assumed to randomly update the store and, in particular  attribute $\loc$.  The computational model is {\em clock-synchronous}, as in \FlyFast{,} but at the component level.
In addition, each component is equipped with a local {\em outbox}. The effect of an output action 
$\alpha^*[\pi_r]\tuple{}\sigma$ is to deliver output label $\alpha\tuple{}$ to the local outbox,
together with the predicate $\pi_r$, which (the store of) the receiver components will be required to satisfy, as
well as the current store  of the component executing the action; the current store
is updated according to update $\sigma$. Note that output actions are {\em non-blocking}
and that successive output actions of the same component overwrite its outbox. 
An input action $\alpha^*[\pi_s]()\sigma$ by a component
will be executed with a probability which is proportional to the {\em fraction} of all those
components whose outboxes currently contain the label $\alpha\tuple{}$, a predicate $\pi_r$
which is satisfied by the component, and a store  which satisfies in turn 
predicate $\pi_s$. If such a fraction is zero, then the input action will not take place (input is blocking),
otherwise the action takes place, the store of the component is updated via $\sigma$, and its outbox
cleared.

\paragraph{Related Work}
CAS are typically {\em large} systems, so that the formal analysis of models for such systems
hits invariantly the state-space explosion problem. In order to mitigate this problem, the so called `on-the-fly' paradigm 
is often adopted (see e.g.~\cite{Courcoubetis1992,BCG95,Hol04,GnM11}).

In the context of probabilistic model-checking several on-the-fly approaches have been proposed, among which~\cite{DIMTV04}, \cite{LLM14a} and~\cite{Ha+09}.  In~\cite{DIMTV04}, a probabilistic model-checker is shown for the
{\em time bounded} fragment of  PCTL.  An on-the-fly approach for {\em full} PCTL model-checking is proposed in~\cite{LLM14a} where, actually, a specific {\em instantiation} is presented of an algorithm  which is {\em parametric} with respect to the specific probabilistic processes modelling language and logic, and their specific semantics.
Finally, in~\cite{Ha+09} an on-the-fly approach is used for detecting a maximal relevant search depth in an infinite state space and then a {\em global} model-checking approach is used
for verifying bounded Continuous Stochastic Logic (CSL)~\cite{Az+00,Ba+03} formulas in a continuous time setting on the selected subset of states. 

An on-the-fly approach by itself however, does not solve the challenging scalability problems that arise in truly large parallel systems, such as CAS.
To address this type of scalability challenges in probabilistic model-checking, recently, several approaches have been proposed. In~\cite{He+04,Gu+06} approximate probabilistic model-checking is introduced. This is a form of statistical model-checking that consists in the generation of random executions of an {\em a priori} established maximal length~\cite{LaL16}. On each execution the property of interest is checked and statistics are performed over the outcomes. The number of executions required for a reliable result depends on the maximal error-margin of interest. The approach relies on the analysis of individual execution traces rather than a full state space exploration and is therefore memory-efficient. However, the number of execution traces that may be required to reach a desired accuracy may be large and therefore time-consuming. The approach works for general models, i.e.
models where stochastic behaviour can also be non Markovian and that do not necessarily model populations of similar objects.
On the other hand,  the approach is not independent from the number of objects involved. 
As recalled above, in~\cite{LLM14} a scalable  model-checking algorithm is presented that is based on mean-field approximation, for the verification of time bounded PCTL properties of an  individual  in the context of a system consisting of a large number of interacting objects.  Correctness of the algorithm with respect to exact probabilistic model-checking has been proven in~\cite{LLM14} as well. Also this algorithm is actually an instantiation of the above mentioned parametric algorithm for (exact) probabilistic model-checking~\cite{LLM14a}, but the algorithm is instantiated on
(time bounded PCTL and) the {\em approximate}, mean-field, semantics of a population process modelling language.
It is worth pointing out that  \FlyFast{} allows users to perform simulations of their system models and to analyse the latter using
their {\em exact} probabilistic semantics and {\em exact} PCTL model-checking. In addition, the tool
provides {\em approximate}  model-checking for bounded PCTL, using the model semantics based on mean-field.  

The work of Latella et al.~\cite{LLM14} is based on mean-field approximation in the {\em discrete time} setting; 
approximated mean-field model-checking in the {\em continuous time} setting has been presented in the literature as well, where 
the deterministic approximation of the global system behaviour is formalised as an initial value problem using a set of { differential}
equations.
Preliminary ideas on the exploitation of mean-field convergence in continuous time for model-checking were informally sketched in~\cite{Ko+12}, but no model-checking algorithms were presented. Follow-up work on the above mentioned approach can be found in~\cite{Ko+13} which relies on earlier results on fluid model-checking by Bortolussi and Hillston~\cite{BoH12b}, later published in~\cite{BoH15}, where a {\em global CSL} model-checking procedure is proposed for the verification of properties of a selection of individuals in a population, which relies on fast simulation results.
This work is perhaps closest related to~\cite{LLM14,LLM15a}; 
however their procedure exploits mean-field convergence and fast 
simulation~\cite{DaN08,GaG10} in a {\em continuous} time setting---using a set of { differential} equations---rather than in a discrete time setting---where an inductive definition 
is used. Moreover, that approach is based on an {\em interleaving} model of computation, rather than
a clock-synchronous one; furthermore, a {\em global } model-checking approach,
rather than an on-the-fly approach is adopted; it is also worth noting that the treatment of nested formulas, 
whose truth value may change over time, 
turns out to be much more difficult in the interleaving, continuous time, global model-checking approach
than in the clock-synchronous, discrete time, on-the-fly one.


\Abf{} has been originally proposed in~\cite{CLM16a}, where the complete formal, exact probabilistic, semantics 
of the language have been defined. The semantics definition consists of three transition rules---one for
transitions associated with output actions, one for those associated with input actions, and one for transitions
to be fired with residual probability. The rules induce a transition relation among component states
and compute the relevant probabilities. From the component transition relation, a component one-step
transition probability matrix is derived, the elements of which may depend on the fractions of the components
in the system which are in a certain state. The system-wide one-step 
transition probability matrix is obtained by product---due to independence assumptions---using the 
above mentioned component probability matrix and the actual fractions in the current system global state.
In~\cite{CLM16a} a translation of \Abf{} to the model specification language
of \FlyFast{}  has also been presented which makes \Abf{} an additional front-end for \FlyFast{} extending its applicability
to models of systems based on predicate-based interaction. In the above mentioned paper, correctness of the translation
has been proved as well. In particular, it has been shown that the probabilistic semantics of any \Abf{} model
are isomorphic to those of the translation of the model. In other words, the transition probability matrix 
of (the DTMCs of) the two models is the same. A companion translation of bounded PCTL formulas
is also defined~\cite{CLM16a} and proven correct.

The notion of the outbox used in \Abf{} is reminiscent of the notion of the {\em ether} in PALOMA~\cite{FeH14} 
in the sense that the collection of all outboxes together can be thought of as a kind
of ether; but such a collection is intrinsically distributed among the components so that
it cannot represent a bottleneck in the execution of the system neither a singularity point in the
deterministic approximation.

We are not aware of other proposals, apart from~\cite{CLM16a}, of probabilistic process languages,  
equipped both with standard, DTMC-based, semantics and with mean-field ones, 
that provide a predicate-based interaction framework, and
that are fully supported by a tool for probabilistic simulation, exact and mean-field model-checking.

We conclude this section recalling that 
mean-field/fluid procedures are based on  {\em approximations} of the
global behaviour of a system. Consequently, the techniques should be considered as  {\em complementary}
to other, possibly more accurate but often not as scalable, analysis techniques for CAS, primarily those based on stochastic simulation, such as
statistical model-checking.

In this paper we present some details of \Abf{,} a translation to \FlyFast{} which simplifies that proposed in~\cite{CLM16a} and an approach to model reduction
based on probabilistic bisimulation for Inhomogeneous DTMCs. In Section~\ref{FFaAbfe} 
we briefly present the main ingredients of the \Abf{} syntax and informal semantics, and 
we recall those features of \FlyFast{} directly relevant for understanding
the translation of  \Abf{} to the  \FlyFast{} input language proposed in~\cite{CLM16a}.
A  revised and simplified version of the translation is described in Section~\ref{Art}.
In Section~\ref{sec:stpLanguage} we introduce a simplified language for the definition of transition-probabilities in \Abf{} that allows us to define in Section~\ref{sec:Bisimilarity} a model reduction procedure of the translation result, based on a notion of bisimulation for the kind of IDTMCs of interest, introduced in Section~\ref{sec:Bisimilarity} as well. An example of application of the 
procedure is presented in Section~\ref{RedEx}. Some conclusions are drawn in Section~\ref{Concl}.
A formal proof of decidability of the cumulative probability test for state-space reduction based on
bisimulation is provided in the Appendix.

\section{Summary on \Abf{} and \FlyFast{}}
\label{FFaAbfe}
In the following we present the main ingredients of \Abf{} and the features of \FlyFast{} relevant for the present paper.

\subsection{\Abf{}}
A \Abf{} system model specification $\Upsilon = (\os_{\Upsilon}, F_{\Upsilon}, \boldsymbol{\Sigma_0})\aN$
is a triple where $F_{\Upsilon}$ is the set of relevant function definitions (e.g. store updates, auxiliary constants and functions),
$\os_{\Upsilon}$ is  a set of state defining equations, and
$\boldsymbol{\Sigma_0}$ is the initial system state (an $N$-tuple
of component states, each of which being a 3-tuple $(C,\gamma,O)$ of agent state $C$,
store $\gamma$ and outbox $O$). 
We describe the relevant details below referring to~\cite{CLM16a} for the formal definition probabilistic semantics
of the language.%

\noindent
The \Abf{}  type system consists of floating point values and operations, as in \FlyFast{},
plus simple enumeration types for attributes, declared according to the syntax
$\attype\; \mathtt{<name>} \; \enum\; \mathtt{<id-list>}$. 
%
$\mathtt{<id-list>}$ is a finite list of identifiers.
Of course, attributes can also take floating point values. 

\begin{figure}
\noindent
{\footnotesize
$
\attype\; \mathtt{Space}\; \enum\; \mathtt{A,B,C,D};\\
\vdots\\
\const\; \mathtt{H} = 0.6;\\
\const\; \mathtt{L = 1- H};\\
\const\; \mathtt{Hdiv2 = H/2};\\
\const\; \mathtt{Ldiv2 = L/2};\\
\vdots\\
\attribute \; \loc : \mathtt{Space};\\
\vdots\\
\func\; \mathtt{Hr}(x:\mathtt{Space}): \mathtt{Space};\;
x\;
\endfunc;\\
\func\; \mathtt{N}(x:\mathtt{Space}): \mathtt{Space};\;
\case \; x \; \of \;  \mathtt{A:A;B:B;C:B;D:A}\;
\endfunc;\\
\func\; \mathtt{S}(x:\mathtt{Space}): \mathtt{Space};\;
\case \; x \; \of \;  \mathtt{A:D;B:C;C:C;D:D}\;
\endfunc;\\
\func\; \mathtt{E}(x:\mathtt{Space}): \mathtt{Space};\;
\case \; x \; \of \;  \mathtt{A:A;B:A;C:D;D:D}\;
\endfunc;\\
\func\; \mathtt{W}(x:\mathtt{Space}): \mathtt{Space};\;
\case \; x \; \of \;  \mathtt{A:B;B:B;C:C;D:C}\;
\endfunc;\\
\vdots\\
\func\; \mathtt{pHr}(x:\mathtt{Space}): \float;\;
\case \; x \; \of \;  \mathtt{A: H;B: L; C:H;D: L}\;
\endfunc;\\
\func\; \mathtt{pN}(x:\mathtt{Space}): \float;\;
\case \; x \; \of \;  \mathtt{A:0;B:0;C:Ldiv2;D:Hdiv2}\;
\endfunc;\\
\func\; \mathtt{pS}(x:\mathtt{Space}): \float;\;
\case \; x \; \of \;  \mathtt{A:Ldiv2;B:Hdiv2;C:0;D:0}\;
\endfunc;\\
\func\; \mathtt{pE}(x:\mathtt{Space}): \float;\;
\case \; x \; \of \;  \mathtt{A:0;B:Hdiv2;C:Ldiv2;D:0}\;
\endfunc;\\
\func\; \mathtt{pW}(x:\mathtt{Space}): \float;\;
\case \; x \; \of \;  \mathtt{A:Ldiv2;B:0;C:0;D:Hdiv2}\;
\endfunc;\\
\vdots\\
\update \; \mathtt{Jump}\\
\my.\loc := \mathtt{Hr}(\my.\loc) \; \with \; \mathtt{pHr}(\my.\loc);\\
\my.\loc := \mathtt{N}(\my.\loc) \; \with \; \mathtt{pN}(\my.\loc);\\
\my.\loc := \mathtt{S}(\my.\loc) \; \with \; \mathtt{pS}(\my.\loc);\\
\my.\loc := \mathtt{E}(\my.\loc) \; \with \; \mathtt{pE}(\my.\loc);\\
\my.\loc := \mathtt{W}(\my.\loc) \; \with \; \mathtt{pW}(\my.\loc)\\
\endupdate
$
}
\caption{A  fragment of $F_{SI}$.\label{exa:F:SEIR}}
\end{figure}

In Figure~\ref{exa:F:SEIR} the attribute type 
$\mathtt{Space}$ is defined that consists of four values $\mathtt{A,B,C,D}$ modelling four locations.
Some auxiliary constants are defined, using  the $\const$ construct inherited from  \FlyFast{:}
$\const \; \mathtt{<name>} \, = \, \mathtt{<value>}$.
%

A \Abf{} store update definition  has the following syntax\footnote{In~\cite{CLM16a} a slightly different syntax for store updates has been used.}:

\noindent
{\small
$
\update{} \; upd\\
\my.a_1:= e_{11}, \ldots , \my.a_k:= e_{k1} \with{} \; p_1;\\
\vdots\\
\my.a_1:= e_{1n}, \ldots , \my.a_k:= e_{kn} \with{} \; p_n\\
\endupdate
$\\
}

\noindent 
where 
$upd$ is the update name (unique within the system model specification),
$a_1, \ldots, a_k$ are the attribute names of the component,
$e_{11}, \ldots, e_{kn}$ and $p_1, \ldots, p_n$  are attribute/store-probability 
expressions respectively, with syntax defined according to the grammars
$
e::= v_a  \; \sep \; c_a \; \sep \; \my.a \; \sep \; fn_a(e_1,\ldots,e_m)
$ and
$
p::= v_p  \; \sep \; c_p \; \sep  \; fn_p(e_1,\ldots,e_m).
$
In the above definition of 
attribute expressions $v_a$ is an attribute value (drawn from finite set $\calV$ of attribute values),
$c_a$ is an attribute constant in $\calV$ defined using the $\const$;
$a \in \SET{a_1, \ldots, a_k}$ is an attribute name
and $fn_a$ is an attribute function defined by the user in $F_{\Upsilon}$, which, 
when applied to attribute expressions 
$e_1,\ldots,e_m$ returns an attribute value;
the syntax for such function definitions $\mathit{afd}$ is given below:\\

\noindent
{\small
$
\mathit{afd} ::= \func \; fn_a (x_1:T1, \ldots, x_m:Tm) :T;  \mathit{afb} \; \endfunc\\
$
$\mathit{afb} ::= e \; \sep
\case \; (x_1, \ldots, x_m) \; \of
(v_{a_{11}}, \ldots, v_{a_{m1}}): e_{1};
(v_{a_{12}}, \ldots, v_{a_{m2}}): e_{2};
 \ldots 
(v_{a_{1k}}, \ldots, v_{a_{mk}}): e_{k}$\\
}

\noindent
where $fn_a$ is the name of the attribute function, $x_1:T1, \ldots, x_m:Tm$ are its parameters and their relative types,
$T$ is the type of the result of $fn_a$; $e$, $e_{i}$ are attribute-expressions and $v_{a_{ij}}$ are attribute-values.

\begin{figure}
\begin{center}
\scriptsize
\begin{minipage}{2in}
$
\begin{array}{l l l }
\texttt{S} & \dfas 
& \frc(\texttt{I}):: \texttt{inf}^* [\bot]\tuple{} \texttt{Jump}. \texttt{I +}\\
&& \frc(\texttt{S}):: \texttt{nsc}^* [\bot]\tuple{}\texttt{Jump}. \texttt{S}\\\\
\texttt{I} & \dfas 
& ii:: \texttt{inf}^*[\bot]\tuple{}\texttt{Jump}. \texttt{I +}\\  
&& ir:: \texttt{rec}^*[\bot]\tuple{}\texttt{Jump}. \texttt{S}\\\\\\\\\\\\\\\\
\end{array}
$
\end{minipage}
\hspace{0.5in}
\resizebox{1.5in}{!}{\includegraphics{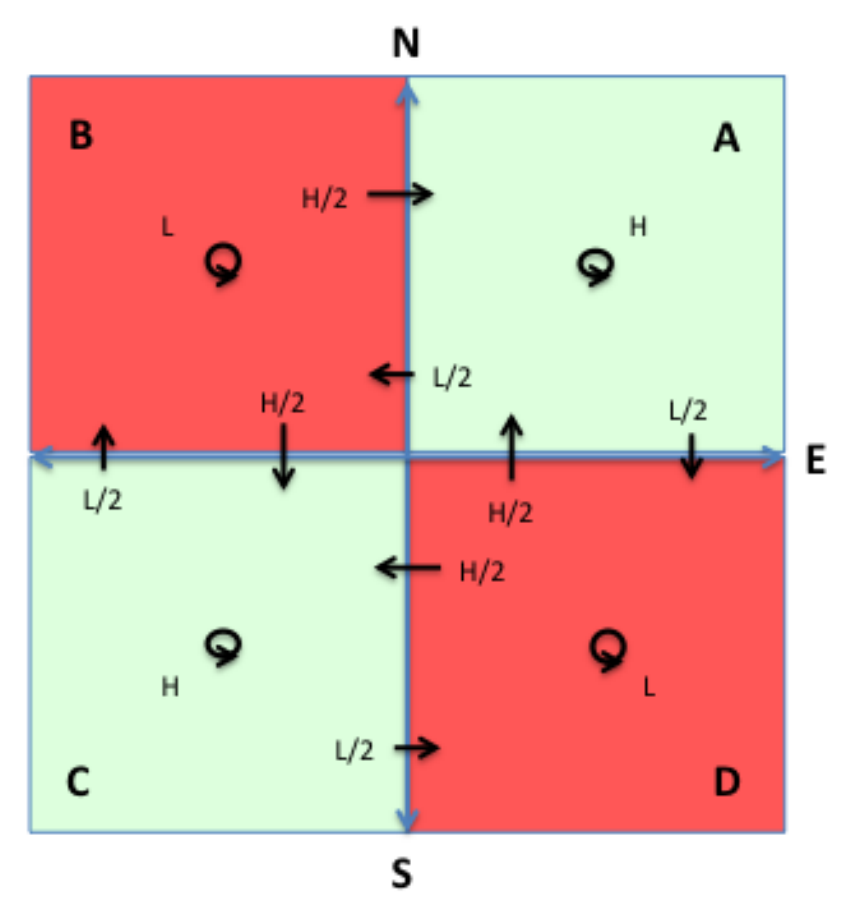}}
\end{center}\vspace{-1in}
\caption{$SI$, a behavioural model.\label{exa:SI}}
\end{figure}

In Figure~\ref{exa:F:SEIR} attribute functions $\mathtt{N,S,E,W}$ are defined for
North, South, East, and West, such that $\mathtt{Space}$ models the Cartesian space with four quadrants: 
$\mathtt{A=N(D)=E(B)}$, $\mathtt{B=N(C)=W(A)}$, and so on, as shown diagrammatically in Figure~\ref{exa:SI} {\em right}. Function $\mathtt{Hr}$ is the identity on $\mathtt{Space}$.
 
In the definition of store-probability expressions $v_p \in (0,1]$, 
$c_p$ is a store-probability constant in $(0,1]$ defined using
the \FlyFast{} $\const$ construct, and $fn_p$ is a store-probability function 
defined by the user in $F_{\Upsilon}$, which, 
when applied to attribute expressions 
$e_1,\ldots,e_m$ returns a probability value.
The syntax for store-probability function definitions $\mathit{pfd}$ is
similar to that of attribute functions:\\

\noindent
{\small
$
\mathit{pfd} ::= \func \; fn_p (x_1:T1, \ldots, x_m:Tm) : \float;  \mathit{pfb} \; \endfunc\\
$
$\mathit{pfb} ::= p \; \sep 
\case \; (x_1, \ldots, x_m) \; \of
(v_{a_{11}}, \ldots, v_{a_{m1}}): p_{1};
(v_{a_{12}}, \ldots, v_{a_{m2}}): p_{2};
 \ldots 
(v_{a_{1k}}, \ldots, v_{a_{mk}}): p_{k}$\\
}

\noindent
where $fn_p$ is the name of the store-probability function, the result type is $\float$ (actually the range $[0,1]$)
$x_1:T1, \ldots, x_m:Tm$ are its parameters and their relative types,
$p$, $p_{i}$ are store-probability expressions and $v_{a_{ij}}$ are attribute-values.
In any store update definition it must be guaranteed that the values of $p_1 \ldots p_n$ sum up\footnote{In this version of the translation we allow only flat updates, i.e. the specific probability of each combination of values assigned
to the attributes must be given explicitly. Other possibilities could be defined using {\em combinations}
of (independent) probability distributions.} to $1$. The informal meaning is clear. The store update will make
attributes $a_1, \ldots, a_k$ take the values of $e_{1i}, \ldots, e_{ki}$ respectively with probability equal to the value of $p_i$.

In Figure~\ref{exa:F:SEIR} store-probability functions $\mathtt{pHr, pN, pS, pE, pW}$ are defined
that give the probabilities of not moving ($\mathtt{pHr}$), or of jumping to North ($\mathtt{pN}$), South  ($\mathtt{pS}$), East  ($\mathtt{pE}$), West  ($\mathtt{pW}$), as functions of the  current location.

\begin{example}
\label{exa}
A simplified version of the behaviour of the epidemic process discussed in~\cite{CLM16a}
is shown in Figure~\ref{exa:SI} {\em left}\footnote{We focus only on those features that are most relevant 
for the present paper. In~\cite{CLM16a} also other features are shown like, e.g. the use of 
(predicate-based) input actions, which are not the main subject of this paper.}.
In Figure~\ref{exa:F:SEIR} we show a fragment of $F_{SI}$ defining store update $\mathtt{Jump}$ together with the relevant type, constant and function definitions as introduced above. The component has just one attribute, named $\loc$, with values in $\mathtt{Space}$. The effect of $\mathtt{Jump}$ executed by a component 
in which $\loc$ is bound to quadrant $\ell$ is to leave the value of  $\loc$ unchanged with probability $\mathtt{pHr}(\ell)$, change it to the quadrant North of $\ell$ with probability  
$\mathtt{pN}(\ell)$, and so on.  Note that $\mathtt{H} > \mathtt{L}$ and this implies that
higher probability is assigned to $\mathtt{A}$ and $\mathtt{C}$ and low probability to $\mathtt{B}$ and $\mathtt{D}$.
This is represented in Figure~\ref{exa:SI} {\em right} where higher probability locations are shown in green and
lower probability ones are shown in red; moreover, the relevant probabilities are represented as arrows ($\mathtt{H/2}, \mathtt{L/2}$)
or self-loops ($\mathtt{H}, \mathtt{L}$).
A susceptible (state $\mathtt{S}$) component becomes infected (state $\mathtt{I}$) via 
an $\mathtt{inf}$ action which takes place with probability equal to the fraction of components in the system which are currently infected (i.e. $\mathtt{frc(I)}$); it remains in state  $\mathtt{S}$ via the self-loop
labelled by action $\mathtt{nsc}$, with probability $\mathtt{frc(S)=1- frc(I)}$. An infected node (state $\mathtt{I}$)
may recover, entering state $\mathtt{S}$ with action $\mathtt{rec}$ and probability $\mathtt{ir}$;
while infected, it keeps executing action $\mathtt{inf}$, with probability $\mathtt{ii}$.
Note that,
for the sake of simplicity, we use only {\em internal} actions, modelled by means of 
{\em output} actions with predicate false ($\bot$). We assume that in the initial global state
all outboxes are non-empty; each contains the initial store of the specific component (i.e., its initial location), predicate $\bot$ and the empty tuple $\tuple{}$).

\end{example}
\noindent
A \Abf{} state defining equation  has the following (abstract) form: 
$
C:= \sum_{j \in J} [g_j]p_j::act_j.C_j 
$
where either $[g_j]p_j$ is the keyword $\rest$ or:
\begin{itemize}
\item $g_j$ is a boolean expression $b$ which may depend on the current store, but not on the
current occupancy measure vector: 
$
b::= \top \; \sep \; \bot \; \sep \; e \, \underline{\bowtie} \, e \; \sep \; \neg b \; \sep \; b \AND b 
$ and 
$
e::= v_a  \; \sep \; c_a \; \sep \my.a
$
where $\top$   ($\bot$) denotes the constant $\mathtt{true}$ ($\mathtt{false}$),
$\underline{\bowtie}\, \in \SET{\geq,>,\leq,<}$, 
$v_a$ is an attribute value (drawn from finite set $\calV$ of attribute values),
$c_a$ is an attribute constant in $\calV$ defined using
the \FlyFast{} $\const$ construct, and $a$ is the name of an attribute of the component.
\item $p_j$ is a transition probability expression 
$
p::= v_p \, \sep   \, c_p \, \sep   \, \frc(C) \, \sep \, \frc(\pi)\, \sep \, \prod_{i \in I} \; p_i \, \sep \, \sum_{i \in I} p_i \, \sep \, 1-p
$, for finite $I$,
where $v_p \in (0,1]$,  $c_p$ a constant in $(0,1]$ defined via the $\const$ construct, 
and $\pi$ is defined as $b$ above, but where expressions $e$ can also be attribute names $a$ 
(i.e. $e::= v_a  \; \sep \; c_a \; \sep \; \my.a \; \sep \; a$);
$\frc(C)$  is the fraction of components {\em currently} in state $C$ over the total number $N$;
similarly, $\frc(\pi)$ is the fraction of components the {\em current} 
store of which satisfies $\pi$, over the total number $N$. Note that it must be guaranteed
that $\prod_{i \in I} p_i \leq 1$  and $\sum_{i \in I} p_i \leq 1$.
\item $act_j$  can be an output action $\alpha^*[\pi]\tuple{}\sigma$ or an input action $\alpha^*[\pi]()\sigma$,
where $\pi$ is as above and $\sigma$ is the 
name of a store update. Note that in the case of an input action, $\pi$ refers to the store
of the partner component in the {\em previous} step of the computation. 
\end{itemize}
If $[g_j]p_j=\rest$, then $act_j$ must be an output action $\alpha^*[\pi]\tuple{}\sigma$, 
to be executed with the residual probability.

\subsection{\FlyFast{}}
\FlyFast{} accepts a specification $\tuple{\os,A,\vct{C_0}}\aN$ of a model of a system consisting of the clock-synchronous product of $N$ instances of a probabilistic agent. The states of the DTMC-like agent model are specified by a set of state-defining equations
$\os$. The (abstract) form of a state defining equation is the following 
$
C:= \sum_{i=1}^r a_i.C_i$
where $a_i \in \act{}$---the set of \FlyFast{} {\em actions}---$C,C_i \in \sc{}$---the set of \FlyFast{} {\em states}---and, for $i,j=1,\ldots,r$ $a_i\not=a_j$ if $i\not=j$; note that $C_i=C_j$ with $i\not=j$ is allowed instead\footnote{The concrete \FlyFast{} syntax is:
{\tt state C$\{$a$\_$1.C$\_1$ + a$\_2$.C$\_2$ \ldots a$\_$r.C$\_$r$\}$}.}.
Each action has a probability assigned by means of 
an action probability function definition in $A$ of the form $a:: exp$ where $exp$ is an expression consisting  of  constants
and  $\frc(C)$ terms. Constants are floating point
values or names associated to such values using the construct  $\const \; \mathtt{<name>} \, = \, \mathtt{<value>}$;
$\frc(C)$ denotes the element associated to state $C$
in the current occupancy measure vector\footnote{The occupancy measure vector is a vector 
with as many elements as the number of states of an individual agent; the
element associated to a specific state gives the fraction of the subpopulation currently in that state over the size of the overall population. The occupancy measure vector is a compact representation of the system global state.}. So, strictly speaking, $\os$ and $A$ characterise an inhomogeneous DTMC whose probability matrix $\otm(\vct{m})$ 
is a function of the occupancy measure vector $\vct{m}$ such that for each pair of states $C,C'$, 
the matrix element $\otm(\vct{m})_{C,C'}$ is the probability of jumping from  $C$ to $C'$ given the current occupancy measure vector $\vct{m}$. Letting $\sc{\os}$ be the set
of states of the agent, with $|\sc{\os}|=S$,
and $\us{S} = \SET{(m_1,\ldots,m_S) | m_1+ \ldots + m_S=1}$ denote the unit simplex of dimension $S$, we have
$\otm: \us{S} \times \sc{\os} \times \sc{\os} \rightarrow [0,1]$. Matrix $\otm$ is generated directly from the 
input specification $\tuple{\os,A,\vct{C_0}}\aN$; the reader interested in the details of how to derive
$\otm$ is referred to~\cite{LLM14,LLM15a}.
Auxiliary function definitions can be specified in $A$. The initial state $\vct{C_0}$ is a vector of size $N$ consisting of
the initial state of each individual object.
Finally, note that in matrix $\otm(\vct{m})$ the information on specific actions is lost, which is  common in PCTL/DTMC based approaches; furthermore, we note that, by construction, $\otm(\vct{m})$ does not depend on $N$ (see~\cite{LLM14,LLM15a}  for details).

\section{A revised translation}
\label{Art}
As in~\cite{CLM16a}, we define a translation  such that, 
given a \Abf{} system specification $\Upsilon = ( \os_{\Upsilon}, F_{\Upsilon}, \boldsymbol{\Sigma_0})\aN$, 
the translation returns the \FlyFast{} system  specification $\tuple{\os,A,\vct{C_0}}\aN$
preserving probabilistic semantics. 
The predicate-based \FlyFast{} front-end
is then completed with a simple translation at the PCTL level, for which we refer to~\cite{CLM16a}.

The system model specification translation consists of two phases.
In the first phase,  each action in the input system model specification $\Upsilon$ is annotated with an identifier
which is unique within the specification. We let $\aleph(\Upsilon)$ denote the resulting specification.
These annotations will make action names unique specification-wide thus eliminating 
complications which may arise from multiple occurrences of the same action, in particular when leading to the same state (see~\cite{CLM16a} for details).
Of course, these annotations are disregarded in the probabilistic semantics,
when considering the interaction model of components. In other words, an
output action  $\alpha\tuple{}$ in  outbox $(\gamma, \pi, \alpha\tuple{})$ must match with any input action 
$\alpha()$ even if $\alpha\tuple{}$ would actually correspond to $(\alpha,\iota)^*[\pi]\tuple{}$ and $\alpha()$
would actually correspond to $(\alpha,\eta)^*[\pi']()$. 
Apart from this detail, the probabilistic semantics as defined in~\cite{CLM16a} remain unchanged.

The second phase is defined by the translation algorithm defined in Figure~\ref{alg:trans},
which is a revised and simplified version of that presented in~\cite{CLM16a} and is applied to $\aleph(\Upsilon)$.
We let $\calI(\aleph(\Upsilon))$ denote the result of the translation, namely the pure \FlyFast{} system
specification $\tuple{\os,A,\vct{C_0}}\aN$.
 %
%

We recall here  some notation from~\cite{CLM16a}. We let $\sc{\os_{\Upsilon}}$ denote the
set of states of $\Upsilon$; $\Gamma_{\os_{\Upsilon}}$ is the set of all stores defined
over the attributes of $\Upsilon$---a store is a finite mapping from the attributes of 
 the component to a finite set of values $\calV$, thus $\Gamma_{\os_{\Upsilon}}$ is finite---and $\calO_{\os_{\Upsilon}}$ the finite set of all outboxes of $\Upsilon$.
A  $\Upsilon$ {\em component-state} is a triple 
$(C,\gamma, O) \in \sc{\os_{\Upsilon}} \times \Gamma_{\os_{\Upsilon}} \times \calO_{\os_{\Upsilon}}=\Omega_{\os_{\Upsilon}}$.
If the component-state is the target of a transition modelling the execution of an {\em output} action,
then $O=(\gamma', \pi, \alpha\tuple{})$, where $\gamma'$ is the store of the (component-state) source  of the transition,
$\pi$ is the predicate used in the action---actualised with $\gamma'$---and $\alpha\tuple{}$ the actual message sent by the action. 
If, instead, the component-state is the target of a transition for an {\em input} 
 action, then $O=\tuple{}$, i.e. the empty outbox. Note that the set of component states of $\aleph(\Upsilon)$ is identical to that of $\Upsilon$. 
 Also the set of all stores of $\aleph(\Upsilon)$ is the same as
 that of $\Upsilon$.  
In the algorithm of Figure~\ref{alg:trans} by $t*t'$ we mean the {\em syntactical} term representing
the product of terms $t$ and $t'$; the notation is extended to $\synprod\SET{t | \cond(t)}$, denoting  the {\em syntactical}  product $t_1 * \ldots * t_n$ 
if $\SET{t|\cond(t) = \true}=\SET{t_1, \ldots, t_n}\not=\emptyset$ 
and $1$ otherwise. Similarly, $\synsum\SET{t | \cond(t)}$ denotes the {\em syntactical}
sum  $t_1 + \ldots + t_n$ if $\SET{t|\cond(t) = \true}=\SET{t_1, \ldots, t_n}\not=\emptyset$ 
and $0$ otherwise.
The translation algorithm uses a few auxiliary functions which we briefly discuss below:
\begin{itemize}
\item $\calI_{\sc{}}: \Omega_{\os_{\Upsilon}} \rightarrow \sc{}$ is a total injection which maps every component state of $\aleph(\Upsilon)$ to a distinct state of  $\calI(\aleph(\Upsilon))$; we recall that $\sc{}$ denotes the set of state names of \FlyFast{} models. 
\item $\calI_{\act{}}: (\sc{\os_{\Upsilon}} \times \Gamma_{\os_{\Upsilon}}) \times (\Lambda_{\os_{\Upsilon}} \times I_{\aleph}) \times  \Omega_{\os_{\Upsilon}} \rightarrow \act{}$ is a total injection where, as in~\cite{CLM16a}, $\Lambda_{\os_{\Upsilon}}$ is the set of action labels of $\Upsilon$ and  
$I_{\aleph}$ is the set of unique identifiers used in the first phase of the translation. We recall that
$\act{}$ is the set of action names of \FlyFast{.}
The mapping of actions is a bit more delicate because we have to respect 
\FlyFast{} static constraints and, in particular, we have to avoid multiple
probability function definitions for the same action. A first source of potential violations (i.e. multiple
syntactical occurrences of the same action) has been removed by action annotation in the first phase of the translation.
A second source is the fact that  the same action can take place in different contexts (for example
with different stores) or leading to different target component states (maybe with different probabilities).
To that purpose, we could distinguish different occurrences of the same action in different transitions, each characterised
by its source component-state and its target component-state in $\Omega_{\os_{\Upsilon}}$. In practice, since
an action of a component cannot be influenced by the current outbox of the component, 
it is sufficient to restrict the first component of the domain from $\Omega_{\os_{\Upsilon}} $ to $(\sc{\os_{\Upsilon}} \times \Gamma_{\os_{\Upsilon}})$.

\item The  interpretation functions defined in Figure~\ref{semint}, namely those depending
on stores only (and not on occupancy measure vectors); we assume $\semint{\mathbf{E_L}}{\cdot}_{\gamma}$ 
extended to  $\semint{\mathbf{E_L}}{fn}_{\gamma}$ for defined function $fn$, in the standard way.
In Figure~\ref{semint} $\beta_{\Upsilon}$ denotes the constant to value bindings 
generated by the $\const$ construct in the input model specification $\Upsilon$, whereas
store update $upd$ is defined as above.
\item The translation function $\calI_{\calP}$ for transition probability expressions $p_j$, defined in Figure~\ref{tprobexp}.
\end{itemize}

\begin{figure}[h!]
$
\begin{array}{l c l}
\semint{\mathbf{E_L}}{\top}_{\gamma} & = &\true\\
\semint{\mathbf{E_L}}{\bot}_{\gamma} & = &\false\\
\semint{\mathbf{E_L}}{e_1 \, \underline{\bowtie} \, e_2}_{\gamma} & = &
\semint{\mathbf{E_L}}{e_1}_{\gamma} \, \underline{\bowtie} \, \semint{\mathbf{E_L}}{e_2}_{\gamma} \\
\semint{\mathbf{E_L}}{\neg b}_{\gamma} & = &\neg \semint{\mathbf{E_L}}{b}_{\gamma}\\
\semint{\mathbf{E_L}}{b_1  \AND  b_2}_{\gamma} & = &
\semint{\mathbf{E_L}}{b_1}_{\gamma}  \AND  \semint{\mathbf{E_L}}{b_2}_{\gamma}\\
\semint{\mathbf{E_L}}{v_a}_{\gamma} & = & v_a\\
\semint{\mathbf{E_L}}{c_a}_{\gamma} & = & \beta_{\Upsilon}(c_a)\\
\semint{\mathbf{E_L}}{v_p}_{\gamma} & = & v_p\\
\semint{\mathbf{E_L}}{c_p}_{\gamma} & = & \beta_{\Upsilon}(c_p)\\
\semint{\mathbf{E_L}}{a}_{\gamma} & = & a\\
\semint{\mathbf{E_L}}{\my.a}_{\gamma} & = & \gamma(a)
\end{array}
$\\
$
\begin{array}{l c l}
\semint{\mathbf{E_L}}{fn_a(e_1,\ldots,e_m)}_{\gamma} & = &
\semint{\mathbf{E_L}}{fn_a}_{\gamma} (\semint{\mathbf{E_L}}{e_1}_{\gamma}, \ldots, \semint{\mathbf{E_L}}{e_m}_{\gamma})\\
\semint{\mathbf{E_L}}{fn_p(e_1,\ldots,e_m)}_{\gamma} & = &
\semint{\mathbf{E_L}}{fn_p}_{\gamma} (\semint{\mathbf{E_L}}{e_1}_{\gamma}, \ldots, \semint{\mathbf{E_L}}{e_m}_{\gamma})
\end{array}
$\\\\
$
\begin{array}{l c l}
\semint{\mathbf{E_U}}{upd}_{\gamma} & = & 
\lambda \gamma' . \mathtt{dom}(\gamma')\not=\SET{a_1,\ldots, a_k} \rightarrow 0;\\
&   & \hspace{0.22in} 
\gamma'(a_1) = \semint{\mathbf{E_L}}{e_{11}}_{\gamma} \AND \ldots \AND \gamma'(a_k) = \semint{\mathbf{E_L}}{e_{k1}}_{\gamma}
\rightarrow \semint{\mathbf{E_L}}{p_1}_{\gamma};\\
&   & \hspace{0.22in}  \vdots\\
&   & \hspace{0.22in} 
\gamma'(a_1) = \semint{\mathbf{E_L}}{e_{1n}}_{\gamma} \AND \ldots \AND \gamma'(a_k) = \semint{\mathbf{E_L}}{e_{kn}}_{\gamma}
\rightarrow \semint{\mathbf{E_L}}{p_n}_{\gamma};\\
&   & \hspace{0.22in} 
\otherwise \; \rightarrow \; 0
\end{array}
$\\\\
$
\begin{array}{l c l}
\semint{\mathbf{E_R}}{\top}_{\gamma} & = &\true\\
\semint{\mathbf{E_R}}{\bot}_{\gamma} & = &\false\\
\semint{\mathbf{E_R}}{e_1 \, \underline{\bowtie} \, e_2}_{\gamma} & = &
\semint{\mathbf{E_R}}{e_1}_{\gamma} \, \underline{\bowtie} \, \semint{\mathbf{E_R}}{e_2}_{\gamma} \\
\semint{\mathbf{E_R}}{\neg b}_{\gamma} & = &\neg \semint{\mathbf{E_R}}{b}_{\gamma}\\
\semint{\mathbf{E_R}}{b_1  \AND  b_2}_{\gamma} & = &
\semint{\mathbf{E_R}}{b_1}_{\gamma}  \AND  \semint{\mathbf{E_R}}{b_2}_{\gamma}\\
\semint{\mathbf{E_R}}{v_a}_{\gamma} & = & v_a\\
\semint{\mathbf{E_R}}{c_a}_{\gamma} & = & \beta_{\Upsilon}(c_a)\\
\semint{\mathbf{E_R}}{a}_{\gamma} & = & \gamma(a)\\
\end{array}
$\\
\caption{\label{semint} Interpretation functions relevant for the translation}
\end{figure}
\begin{figure}
$
\begin{array}{l c l}
\calI_{\calP}(v_p)_{\gamma} & = & v_p\\
\calI_{\calP}(c_p)_{\gamma} & = & \beta_{\Upsilon}(c_p)\\
\calI_{\calP}(\frc(C))_{\gamma} & = & 
\synsum\SET{\frc (\calI_{\sc{}}((C',\gamma',O')))\;|\;(C',\gamma',O')  \in  \Omega_{\os_{\Upsilon}} \mbox{ and } C' = C}\\
\calI_{\calP}(\frc(\pi))_{\gamma} & = & 
\synsum\SET{\frc (\calI_{\sc{}}((C', \gamma',O')))\; |\; (C',\gamma',O')  \in  \Omega_{\os_{\Upsilon}} \mbox{ and }  
\semint{\mathbf{E_R}}{\semint{\mathbf{E_L}}{\pi}_{\gamma}}_{\gamma'}=\true}\\
\calI_{\calP}(\prod_{i \in I} \; p_i)_{\gamma} & = &  \synprod\SET{\calI_{\calP} (p_i)_{\gamma} \; | \; i \in I}\\
\calI_{\calP}(\sum_{i \in I} \; p_i)_{\gamma} & = &  \synsum\SET{\calI_{\calP} (p_i)_{\gamma} \; | \; i \in I}
\end{array}
$
\caption{\label{tprobexp} Transition probability expressions translation function definition}
\end{figure}

\begin{figure}[!h]
\fbox{
\parbox{6.1in}
{
\scriptsize
\noindent
For each   state equation
$
C:= \sum_{j \in J} [g_j]p_j::act_j.C_j 
$
in $ \os_{\Upsilon}$:
\begin{enumerate}
\item\label{T:OUT} 
For each {\em output} action $(\alpha,\iota)^*[\pi]\tuple{}\sigma=act_k$ with $k \in J$ and $[g_k]p_k\not=\rest$,\\
for each $\gamma \in \Gamma_{\os_{\Upsilon}}$ 
s.t. $\semint{\mathbf{E_L}}{g_k}_{\gamma}=\true$ 
and $(C,\gamma, O) \in \Omega_{\os_{\Upsilon}}$ for some $O \in  \calO_{\os_{\Upsilon}}$,
for each $\gamma' \in \Gamma_{\os_{\Upsilon}}$ s.t.
$(C_k,\gamma',(\gamma, \semint{\mathbf{E_L}}{\pi}_{\gamma}, \alpha\tuple{})) \in \Omega_{\os_{\Upsilon}}$ and $\semint{\mathbf{E_U}}{\sigma}_{\gamma}(\gamma') >0$,
let 
$\xi = \calI_{\act{}}((C,\gamma),(\alpha\tuple{},\iota),(C_k,\gamma',(\gamma,\semint{\mathbf{E_L}}{\pi}_{\gamma},\alpha\tuple{})))$ be a fresh new action in the \FlyFast{} model specification 
$\calI(\aleph(\Upsilon))= \tuple{\os,A,\vct{C_0}}\aN$ and add the following action probability function definition in $A$:
$\xi::\semint{\mathbf{E_U}}{\sigma}_{\gamma}(\gamma') * \calI_{\calP}(p_k)_{\gamma}$.

Moreover, for each outbox $O \in  \calO_{\os_{\Upsilon}}$ s.t. 
$(C,\gamma, O) \in \Omega_{\os_{\Upsilon}}$, the following summand is
added to the equation in $\os$ for state $\calI_{\sc{}}((C,\gamma,O))$:
$
\xi. \,
\calI_{\sc{}}((C_k,\gamma',(\gamma,\semint{\mathbf{E_L}}{\pi}_{\gamma},\alpha\tuple{})));
$

\item\label{T:IN} 
For each {\em input} action  $(\alpha,\iota)^*[\pi]()\sigma = act_k$, with $k \in J$ and $[g_k]p_k\not=\rest$,\\
for each $\gamma \in \Gamma_{\os_{\Upsilon}}$
s.t. $\semint{\mathbf{E_L}}{g_k}_{\gamma}=\true$ 
and $(C,\gamma, O) \in \Omega_{\os_{\Upsilon}}$ for some $O \in  \calO_{\os_{\Upsilon}}$,
for each $\gamma' \in \Gamma_{\os_{\Upsilon}}$ s.t.
$(C_k,\gamma',\tuple{}) \in \Omega_{\os_{\Upsilon}}$ and $\semint{\mathbf{E_U}}{\sigma}_{\gamma}(\gamma') >0$,\\
let 
$\xi = \calI_{\act{}}((C,\gamma), (\alpha(),\iota),(C_k,\gamma',\tuple{}))$,
be a fresh new action in the \FlyFast{} model specification 
$\calI(\aleph(\Upsilon))= \tuple{\os,A,\vct{C_0}}\aN$ and add the following action probability function definition in $A$:\\
$\xi:: \semint{\mathbf{E_U}}{\sigma}_{\gamma}(\gamma')* \calI_{\calP}(p_k)_{\gamma}*$\\
\mbox{ } \hspace{1in}$*\synsum\{\frc (\calI_{\sc{}}(\Sigma))| \Sigma = (C'', \gamma'',(\overline{\gamma},\overline{\pi},\alpha\tuple{}))
 \in   \Omega_{\os_{\Upsilon}} \AND$\\
\mbox{ } \hspace{2.2in}$ \AND \semint{\mathbf{E_R}}{\overline{\pi}}_{\gamma}=\semint{\mathbf{E_R}}{\semint{\mathbf{E_L}}{\pi}_{\gamma}}_{\overline{\gamma}}=\true\}$.\\
%
%
%
Moreover, for each outbox $O \in  \calO_{\os_{\Upsilon}}$ s.t. 
$(C,\gamma, O) \in \Omega_{\os_{\Upsilon}}$, the following summand is
added to the equation in $\os$ for state $\calI_{\sc{}}((C,\gamma,O))$:
$
\xi. \,
\calI_{\sc{}}((C_k,\gamma',\tuple{}));
$\medskip
%
\item\label{T:RESTOUT}
If there exists $k\in J$ s.t. $[g_k]p_k = \rest$, and $act_k=(\alpha,\iota)^*[\pi]\tuple{}\sigma$,
for each $\gamma \in \Gamma_{\os_{\Upsilon}}$
s.t. $(C,\gamma, O) \in \Omega_{\os_{\Upsilon}}$ for some $O \in  \calO_{\os_{\Upsilon}}$,
let  $A_{\gamma}$ be the set of probability function definitions which has been 
constructed in steps (\ref{T:OUT}) and (\ref{T:IN}) above.  
Let $q_{\gamma}$ be defined by $q_{\gamma} = (1 - \synsum\SET{q | \zeta :: r * q \in A_{\gamma}})$.
%
For all $\gamma' \in \Gamma_{\os_{\Upsilon}}$ s.t. 
$(C_k,\gamma',(\gamma,\semint{\mathbf{E_L}}{\pi}_{\gamma},\alpha\tuple{})) \in \Omega_{\os_{\Upsilon}}$,
let 
$\xi = 
\calI_{\act{}}((C,\gamma), (\alpha,\iota)\tuple{},(C_k,\gamma',(\gamma,\semint{\mathbf{E_L}}{\pi}_{\gamma},\alpha\tuple{})))\in \Omega_{\os_{\Upsilon}}$, 
be a fresh new action in the \FlyFast{} model specification 
$\calI(\aleph(\Upsilon))= \tuple{\os,A,\vct{C_0}}\aN$ and add
the following action probability function definition in $A$:
$\xi :: \semint{\mathbf{E_U}}{\sigma}_{\gamma}(\gamma')* q_{\gamma}$.\\
Moreover, for each outbox $O \in  \calO_{\os_{\Upsilon}}$ s.t. 
$(C,\gamma, O) \in \Omega_{\os_{\Upsilon}}$, the following summand is
added to the equation in $\os$ for state $\calI_{\sc{}}((C,\gamma,O))$:
$
\xi.\,
\calI_{\sc{}}((C_k,\gamma',(\gamma,\semint{\mathbf{E_L}}{\pi}_{\gamma},\alpha\tuple{})));
$\medskip
\item No other action probability function definition and transition is included and
the initial state $\vct{C_0}$ of  $\calI(\Upsilon)$ is defined as 
$\vct{C_0}=\calI_{\sc{}}(\boldsymbol{\Sigma_0})$.
\end{enumerate}
}
$ $
}
\caption{The translation algorithm\label{alg:trans}}
\end{figure}

\noindent
Output actions are dealt with in step $1$  of the algorithm of Figure~\ref{alg:trans}. 
Let us consider,  for example, 
$(\texttt{inf},1)^*[\bot]\tuple{}\texttt{Jump}$ in the definition of state {\tt S}
in Figure~\ref{exa:SI} (assuming annotations are integer values and the action has been annotated with $1$). We know that the possible values for locations are $\texttt{A,B,C,D}$,
so that the set of all stores is  $\SET{\loc} \rightarrow \SET{\texttt{A,B,C,D}}$. The algorithm
generates 12 actions\footnote{Diagonal jumps are not contemplated in the model; technically this comes from the
actual probability values used in the definition of $\texttt{Jump}$. 
}. Let us focus on the action $\xi$ associated to local position $\texttt{A}$ (i.e. $\gamma= [\loc \mapsto \texttt{A}]$) and possible next position $B$ (i.e. $\gamma'= [\loc \mapsto B]$); the algorithm will generate the \FlyFast{} probability function definition  $\xi :: \mathtt{pW}(A)*{(\frc(\mathtt{I1})+\ldots+\frc(\mathtt{In}))}$\footnote{Here we assume that 
$\calI_{\sc{}}(\SET{(C,\gamma,O)  \in  \Omega_{\os_{\Upsilon}}| \,C={\tt I}})=\SET{\mathtt{I1},\ldots,\mathtt{In} } \subset \calS$.} as well as a transition leading to (a state which is the  encoding, via $\calI_{\sc{}}$, of) the component state with ${\tt I}$ as (proper) state, store $\gamma'$, and
outbox $(\gamma, \bot, \texttt{inf}\tuple{})$. Since the action is not depending on the current outbox,
in practice a copy of such a transition is generated {\em for each}  component state
sharing the same proper state ${\tt S}$ and the same store $\gamma$.   
The translation scheme for input actions is defined in case $2$ and is similar, except that
one has also to consider the sum  of the fractions of the possible partners. 
The translation of the $\rest$ case is straightforward. Note that for
every $\zeta :: r * q \in A_{\gamma}$, $r$ is a probability value associated to a store update;
since any store update characterizes a probability distribution
over stores, assuming the range of such a distribution is
$\SET{r_1,\ldots,r_n}$ if $\zeta_i :: r_i * q \in A_{\gamma}$, then also 
$\zeta_j :: r_j * q \in A_{\gamma}$ for all $j=1,\ldots,n$, $j\not=i$ with
$\sum_{i=1}^n r_j = 1$. Thus  the remaining probability is
$q_{\gamma} = (1 - \synsum\SET{q | \zeta :: r * q \in A_{\gamma}})$, where $q$ is either a 
term $\calI_{\calP}(p_j)_{\gamma}$, with $p_j$ occurring in a summand of the state defining equation (see step \ref{T:OUT}), or a term $\calI_{\calP}(p_j)_{\gamma} * \synsum\SET{ \frc (\calI_{\sc{}}(\Sigma)) | \ldots }$ (see step \ref{T:IN}).
It worth pointing out here that the translation of Figure~\ref{alg:trans} is essentially the same as    that
presented in~\cite{CLM16a}, when the latter is applied to the sublanguage of \Abf{} where one requires
that each action occurs at most once. The annotations performed in the 
first phase of the translation ensure that this requirement is fulfilled;
as we noted above, these annotations are purely syntactical and are disregarded at the semantics level.
We also recall that in probabilistic, pure DTMC process language semantics, actions are in the end dropped and,
for each pair of states, the cumulative probability of such actions is assigned to the single transition from
one of the states to the other one. Consequently, correctness of the translation, proved in~\cite{CLM16a},
is preserved by the simplified version presented in this paper.

We note that in the algorithm sets $\Omega_{\os_{\Upsilon}}$, $ \Gamma_{\os_{\Upsilon}}$ and 
$\calO_{\os_{\Upsilon}}$ are used.
Of course, an alternative approach could be one which considers only the set
$\overline{\Omega_{\os_{\Upsilon}}}$ of component states which are {\em reachable}
from a given initial component state and, consequently, the sets 
$\overline{\Gamma_{\os_{\Upsilon}}}$ and  $\overline{\calO_{\os_{\Upsilon}}}$
of {\em used} stores and outboxes. In this way, the size of the resulting \FlyFast{}
model specification would be smaller (for example in terms of number of states).
On the other hand, this approach might require recompilation for each model-checking session
starting from a different initial component state.

\section{A simplified language for Bisimulation-based optimisation}
\label{sec:stpLanguage}

In this section we consider a simplified language for transition probability expressions appearing in 
state defining equations that will allow us to perform 
bisimulation based optimisation of the result $\tuple{\os,A,\vct{C_0}}\aN$. 
%
The restricted syntax for transition probability expressions $p$ we use in this section is the following:
$
p::= e_p \, \sep   \, e_p \cdot\frc(C)\, \sep \, e_p \cdot\frc(\pi)
$ and
$
e_p ::= v_p \, \sep   \, c_p
$
where $v_p$ and $c_p$ and $\pi$ are defined as in Section~\ref{FFaAbfe}.
%

By inspection of the \FlyFast{} translation as defined in Section~\ref{Art}, 
and recalling that the set $\sc{\os}$ of the states of the resulting \FlyFast{} model,
ranged over by $z, z_i, \ldots$, has cardinality $S$,
it is easy to see that  the probability action definition in the result of a translation of a generic {\em output} action 
is either of the form $\xi :: k$,  or it is of the form
$\xi :: k*\synsum\SET{\frc(z_i) | i \in I}$ where  $k$ is a \FlyFast{} constant. Moreover, if $p$ was of the form $e_p \cdot\frc(C)$, then
index set $I \subseteq \SET{1,\ldots, S}$ identifies those states in $\sc{\os}$
that represent (via $\calI_{\sc{}}$) component states with proper local state $C$; 
if instead, $p$ was of the form $e_p \cdot\frc(\pi)$, then $I \subseteq \SET{1,\ldots, S}$ 
identifies those states in $\sc{\os}$ that represent (via $\calI_{\sc{}}$) component states 
with a store satisfying $\pi$ in the relevant store.
At the \FlyFast{} semantics level, recalling that $\frc(z_i)$ is exactly
the $i$-th component $m_i$ of the occupancy measure vector $\vct{m}=(m_1, \ldots, m_S)$
of the model, we can rewrite\footnote{With a little notational abuse
using $k$ also as the actual value in $[0,1]$ of the \FlyFast{} constant $k$.} the above as $k$ or $k\cdot \sum_{i \in I}m_i$.


Similarly, the probability action definition in the result of a translation of a generic {\em input} action $(\alpha,\iota)^*[\pi']()$ 
(executed in local store $\gamma'$)
will necessarily  be of the form
$k\cdot \left(\sum_{j \in I'}  m_j\right)$
or of the form $k\cdot\left(\sum_{j \in I}  m_j\right)\cdot \left(\sum_{j \in I'}  m_j\right)$,
for index sets $I$ as above and $I'$ as follows:\\

$
I'=\{i \in \SET{1,\ldots, S} | \exists C, \gamma, \overline{\gamma}, \overline{\pi}, \, s.t.
$\\
\mbox{ }\hspace{0.7in}
$
z_i = \calI_{\sc{}}((C,\gamma,(\overline{\gamma},\overline{\pi},\alpha\tuple{})))\AND
\semint{\mathbf{E_R}}{\overline{\pi}}_{\gamma'}=
\semint{\mathbf{E_R}}{\semint{\mathbf{E_L}}{\pi'}_{\gamma'}}_{\overline{\gamma}}=\true
\}\\
$

An immediate consequence of using the above mentioned restricted syntax for the probability function definitions
is that, letting  
$\otm: \us{S} \times \sc{} \times \sc{} \rightarrow [0,1]$ 
be the transition probability matrix for the \FlyFast{} translation of a model specification,
we have that $\otm(m_1,\ldots,m_S)_{z,z'}$ is a polynomial function of degree at most $2$
in variables $m_1, \ldots, m_S$.

\section{Bisimilarity and State-space Reduction}
\label{sec:Bisimilarity}
The following definition generalises standard probabilistic bisimilarity for state labelled DTMCs to the case in which transition probabilities are {\em functions} instead of constant values. 

\begin{definition}
\label{bis}
For finite set of states $\sc{}$, with $|\sc{}|=S$, let $\otm: \us{S} \times \sc{} \times \sc{} \rightarrow [0,1]$  and,
for $z \in \sc{} $ and $Q \subseteq \sc{}$, write 
$\otm(\vct{m})_{z,Q}$ for $\sum_{z' \in Q}\otm(\vct{m})_{z,z'}$.
Let furthermore $\calL:\sc{} \rightarrow 2^{AP}$ be a state-labelling function,
for a given set $AP$ of atomic propositions. 
An equivalence relation  $R \subseteq \sc{} \times \sc{}$ 
is called a {\em bisimulation relation} if and only if
$z_1\, R \; z_2$ implies: (i) $\calL(z_1) = \calL(z_2)$ and (ii) $\otm(\vct{m})_{z_1,Q} = \otm(\vct{m})_{z_2,Q}$,
for all $\vct{m} \in \us{S}$ and $Q \in \sc{}/R$. The {\em bisimulation equivalence} on 
$\sc{}$ is the largest  bisimulation relation $R \subseteq \sc{} \times \sc{}$.
\end{definition}
We point out that the notion of bisimilarity does {\em not}
introduce any approximation, and consequently error, in a model and related analyses.  Bisimilarity is
only a way for abstracting from details that are irrelevant for the specific analyses of interest. In particular, 
it  is useful to remark that bisimilarity preserves also state labels, which are directly related 
to the atomic propositions of logic formulas for which model-checking is performed. Actually, 
it is well known that probabilistic bisimilarity coincides with PCTL equivalence, i.e. the equivalence induced on system states
by their satisfaction of PCTL formulas, for finitely branching systems~\cite{BaK08}.

Note that  $\otm(m_1,\ldots,m_S)_{z_1,Q} = \otm(m_1,\ldots,m_S)_{z_2,Q}$ 
for all $(m_1,\ldots,m_S) \in \us{S}$ is in general not decidable. If instead we consider 
only  transition probability matrices as in Section~\ref{sec:stpLanguage},
we see that each side of the above equality is a polynomial function of degree at most $2$
in variables $m_1, \ldots, m_S$ and 
one can define a normal form for 
the polynomial expressions in $m_1, \ldots, m_S$ supported by an ordering relation
on the variable names (e.g. $m_1\prec \ldots \prec m_S$) and get expressions of the general form
$
\left(\sum_{i=1}^{S} \sum_{j \geq i}^{S} h_{ij}\cdot m_i\cdot m_j\right) +
\left(\sum_{i=1}^{S}h_{i}\cdot m_i \right) +
h
$
for suitable $h_{ij},h_{i},h$. Actually, such expressions can always be rewritten
in the form
$
\left(\sum_{i=1}^{S} \sum_{j \geq i}^{S} u_{ij}\cdot m_i\cdot m_j\right) + u
$
for suitable $u_{ij},u$ since, recalling that $\sum_{i=1}^{S} m_i = 1$, we get
$
\sum_{i=1}^{S}h_{i}\cdot m_i = 
\left(\sum_{i=1}^{S} m_i\right)\cdot \left(\sum_{i=1}^{S}h_{i}\cdot m_i\right)
$ 
which, by simple algebraic  manipulation, yields an expression of the following form:
$
\left(\sum_{i=1}^{S} \sum_{j \geq i}^{S} u'_{ij}\cdot m_i\cdot m_j\right)
$; finally, we get 
$
\left(\sum_{i=1}^{S} \sum_{j \geq i}^{S} u_{ij}\cdot m_i\cdot m_j\right) + u
$
by letting $u_{ij} = h_{ij} + u'_{ij}$ and $u=h$.
The following proposition thus establishes decidability of 
$\otm(m_1,\ldots,m_S)_{z_1,Q} = \otm(m_1,\ldots,m_S)_{z_2,Q}$ 
for all $(m_1,\ldots,m_S) \in \us{S}$ for  transition probability matrices as in Section~\ref{sec:stpLanguage}:

\begin{proposition}\label{prop1}$ $\\
Let 
$
A(m_1,\ldots,m_S)= 
\left(\sum_{i=1}^{S} \sum_{j \geq i}^{S} a_{ij}\cdot m_i\cdot m_j\right) +
a
$
and
$
B(m_1,\ldots,m_S)= 
\left(\sum_{i=1}^{S} \sum_{j \geq i}^{S} b_{ij}\cdot m_i\cdot m_j\right) +
b
$
with $a_{ij}, b_{ij}, 
a, b \in \reals$,                 
where $m_1,\ldots,m_S$ are variables taking values over $\reals_{\geq 0}$  
with $\sum_{i=1}^S m_i = 1$. 
The following holds:
$
(\forall m_1,\ldots,m_S. A(m_1,\ldots,m_S) = B(m_1,\ldots,m_S))
\Leftrightarrow
(
(\forall i,j = 1, \ldots, S$ with $i \leq j. a_{ij} = b_{ij}) \wedge
a=b
).
$
\end{proposition}

The above results can be used for reduction of the state-space  of the individual agent, i.e.
the resulting \FlyFast{} model specification, after the application of the
translation described in Section~\ref{Art}, by using for instance the 
standard probabilistic relational coarsest set partition problem algorithm
(see e.g. \cite{HuT92}, page 227) 
with slight obvious 
modifications due to the presence of state-labels and the need of symbolic computation capabilities required for checking (degree 2) polynomial expressions 
equality.\footnote{For instance, on page 227 of~\cite{HuT92}, line 12,  
$v(x,S) = v(y,S)$ should be replaced with $L(x) = L(y) \wedge v(x,S) = v(y,S)$ and
in line 13, $v(x,S) \not= v(y,S)$ should be replaced with $L(x) \not= L(y) \vee v(x,S) \not= v(y,S),$ in order to take state labels into consideration as well. Of course $v(x,S)$ ($L$, respectively)is to be intended as $\otm(\vct{m})_{z,Q}$ ($\calL$, respectively), using the notation  we introduced above for Bisimilarity.
}
It is worth mentioning that state aggregation via bisimilarity is effective only if there is
some sort of compatibility between (i) state labelling---and, consequently, the specific PCTL atomic propositions one uses---and (ii) the way probabilities are assigned to transitions---and, consequently, the cumulative probabilities to equivalence classes. We will come back on this issue in the following section.

\section{Example}
\label{RedEx}
\begin{figure}
\centerline{
\resizebox{1.5in}{!}{\includegraphics{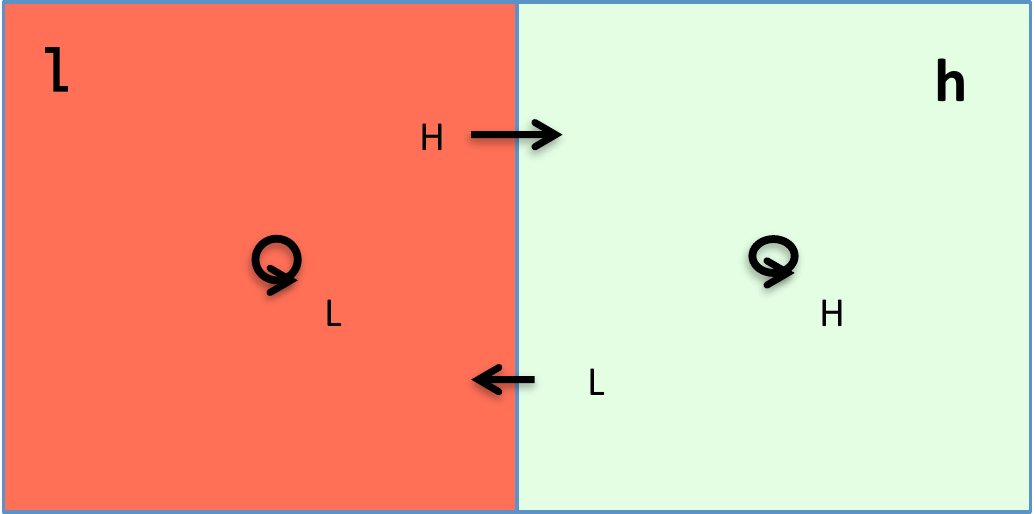}}
}
\caption{$SI$ in two quadrants \label{pic:TwoQuad}}
\end{figure}
%
%
%
%
%
%

The application of the translation to the specification of Example~\ref{exa} generates an agent model with 8 states, say
$\sc{\os} = \SET{SA, SB, SC, SD, IA, IB, IC, ID}$\footnote{Actually the agent resulting from the
translation of Figure~\ref{alg:trans} has a higher number of states due to the different 
possibilities for outbox values. Many of these states are unreachable from the initial 
state since the agent has no input action and we assume an initial unreachable state pruning has been performed.} with associated IDTMC probability transition matrix as 
shown in Figure~\ref{ex:PTM} where
$m_{xy}$ represents the fraction of objects currently in state $xy$ for $x\in \SET{S,I}$
and $y\in \SET{A,B,C,D}$---i.e. the components in state $x$ and  with $\loc=y$ in the original 
specification of Fig~\ref{exa:SI}---so that
$\mathbf{m}=(m_{SA}, m_{SB}, m_{SC}, m_{SD}, m_{IA}, m_{IB}, m_{IC}, m_{ID})$ is the
occupancy measure vector. In Figure~\ref{ex:PTM}, functions $\phi_S$ and $\phi_I$ are used as abbreviations in the obvious way: 
$\phi_S(\mathbf{m}) = m_{SA} + m_{SB} + m_{SC} + m_{SD}$ and
$\phi_I(\mathbf{m}) = m_{IA} + m_{IB} + m_{IC} + m_{ID}$. Let us assume now that
we are interested in checking PCTL formulas on the model of Fig~\ref{exa:SI} which distinguish
components located  in $A$ or $C$ from those located in $B$ or $D$, and those
in state $S$ from those in state $I$, that is we consider
atomic propositions $Sh,Ih,Sl$ and $Il$ and a labelling $\calL$ such that
$\calL(SA)=\calL(SC)=\SET{Sh}$, $\calL(IA)=\calL(IC)=\SET{Ih}$,
$\calL(SB)=\calL(SD)=\SET{Sl}$, and $\calL(IB)=\calL(ID)=\SET{Il}$. 
\def\pS{\phi_S(\mathbf{m})}
\def\pI{\phi_I(\mathbf{m})}
\begin{figure}
\scriptsize
$$
\begin{array}{ c | c | c | c | c | c | c | c | c |}
&   SA & SB & SC & SD  & IA & IB & IC & ID\\\hline
SA & H\pS & \frac{L}{2}\pS & 0 & \frac{L}{2}\pS &H\pI & \frac{L}{2}\pI & 0 & \frac{L}{2}\pI\\\hline
SB &\frac{H}{2}\pS & L\pS & \frac{H}{2}\pS & 0 & \frac{H}{2}\pI & L\pI & \frac{H}{2}\pI &0\\\hline
SC & 0 & \frac{L}{2}\pS & H\pS & \frac{L}{2}\pS & 0 & \frac{L}{2}\pI & H\pI & \frac{L}{2}\pI\\\hline
SD & \frac{H}{2}\pS & 0 & \frac{H}{2}\pS & L\pS & \frac{H}{2}\pI & 0 & \frac{H}{2}\pI & L\pI \\\hline
IA & H ir & \frac{L}{2} ir & 0 & \frac{L}{2} ir & H ii & \frac{L}{2} ii & 0 & \frac{L}{2} ii \\\hline 
IB & \frac{H}{2} ir & L ir  & \frac{H}{2} ir & 0  & \frac{H}{2} ii & L ii  & \frac{H}{2} ii & 0\\\hline
IC & 0 & \frac{L}{2} ir & H ir & \frac{L}{2} ir  & 0 & \frac{L}{2} ii & H ii & \frac{L}{2} ii\\\hline
ID & \frac{H}{2} ir & 0 & \frac{H}{2} ir & L ir & \frac{H}{2} ii & 0 & \frac{H}{2} ii & L ii \\\hline
\end{array}
$$
\caption{IDTMC transition probability matrix $\otm(\vct{m})$, for $\vct{m}$  in $\us{8}$.\label{ex:PTM}}
\end{figure}

Consider relation $R$ on $\sc{\os}$ defined as
$
R = I_{{\sc{\os}}} \cup \SET{(SA,SC),(SB,SD),(IA,IC),(IB,ID)} \cup \SET{(SC,SA),\\(SD,SB),(IC,IA),(ID,IB)}
$
where $I_{{\sc{\os}}}$ is the identity relation on $\sc{\os}$.
It is very easy to show that $R$ is a bisimulation according to Definition~\ref{bis}. Clearly $R$ is an equivalence relation and 
its quotient $\sc{\os}/R$ is the set $\SET{Q_{Sh},Q_{Sl},Q_{Ih},Q_{Il}}$ with 
$Q_{Sh}=\SET{SA,SC},Q_{Sl}= \SET{SB,SD},Q_{Ih}=\SET{IA,IC},Q_{Il}=\SET{IB,ID}$.
In addition, for all $z_1, z_2 \in \sc{\os}$, whenever  $z_1 \, R \,  z_2$, we have
$\calL(z_1)=\calL(z_2)$ and $\otm(\vct{m})_{z_1,Q}=\otm(\vct{m})_{z_2,Q}$ for
all $Q \in \sc{\os}/R$ and for all $\vct{m}$, as one can easily check; clearly, $R$ is also
the largest bisimulation relation on $\sc{\os}$.
The relationship between the two occupancy measure vectors is:
$m_{{Q_{Sh}}} = m_{SA} + m_{SC}$,
$m_{{Q_{Sl}}} = m_{SB} + m_{SD}$,
$m_{{Q_{Ih}}} = m_{IA} + m_{IC}$, and
$m_{{Q_{Il}}} = m_{IB} + m_{ID}$.
%
We can thus use the reduced IDTMC defined by matrix $\widehat{\otm}(m_{{Q_{Sh}}},m_{{Q_{Sl}}},m_{{Q_{Ih}}},m_{{Q_{Il}}})$
shown in Figure~\ref{ex:RPTM}. It corresponds to the \FlyFast{} agent specification $\widehat{\os}$ given in Figure~\ref{RedFFS}. In a sense, the high probability locations $A$ and $C$, in the new model,  
have collapsed into  a single one, namely $h$ and the low probability ones ($B$ and $D$) have
collapsed into $l$, as shown in Figure~\ref{pic:TwoQuad}. We point out again the correspondence
between the symmetry in the space jump probability on one side and the definition of the state labelling function on the other side. Finally, note that a coarser labelling like, e.g. 
$\calL'(SA)=\calL'(SC)=\calL'(IA)=\calL'(IC)=\SET{h}$,
$\calL'(SB)=\calL'(SD)=\calL'(IB)=\calL'(ID)=\SET{l}$
would make the model collapse into one with only two states, $Q_h$ and $Q_l$,
with probabilities $H:Q_h \rightarrow Q_h, H:Q_l \rightarrow Q_h$,
$L:Q_h \rightarrow Q_l$ and $L:Q_h \rightarrow Q_l$ where only the location would be 
modelled whereas information on the infection status would be lost.

\begin{figure}
\scriptsize
\begin{verbatim}
action QSh_inf_QIh: H*(frc(QIh)+frc(QIl));		        action QSh_nsc_QSh: H*(frc(QSh)+frc(QSl));
action QSh_inf_QIl: L*(frc(QIh)+frc(QIl));		        action QSh_nsc_QSl: L*(frc(QSh)+frc(QSl));
action QSl_inf_QIh: H*(frc(QIh)+frc(QIl));		        action QSl_nsc_QSh: H*(frc(QSh)+frc(QSl));
action QSl_inf_QIl: L*(frc(QIh)+frc(QIl));		        action QSl_nsc_QSl: L*(frc(QSh)+frc(QSl));

action QIh_inf_QIh: H*ii;                          action QIh_rec_QSh: H*ir;
action QIh_inf_QIl: L*ii;                          action QIh_rec_QSl: L*ir;
action QIl_inf_QIh: H*ii;                          action QIl_rec_QSh: H*ir;
action QIl_inf_QIl: L*ii;                          action QIl_rec_QSl: L*ir;

state QSh{QSh_inf_QIh.QIh + QSh_inf_QIl.QIl + QSh_nsc_QSh.QSh + QSh_nsc_QSl.QSl}
state QSl{QSl_inf_QIh.QIh + QSl_inf_QIl.QIl + QSl_nsc_QSh.QSh + QSl_nsc_QSl.QSl}
state QIh{QIh_inf_QIh.QIh + QIh_inf_QIl.QIl + QIh_rec_QSh.QSh + QIh_rec_QSl.QSl}
state QIl{QIl_inf_QIh.QIh + QIl_inf_QIl.QIl + QIl_rec_QSh.QSh +QIl_rec_QSl.QSl}
\end{verbatim}
\caption{Reduced agent specification $\widehat{\os{}}$\label{RedFFS}}
\end{figure}

\begin{figure}
\scriptsize
$$
\begin{array}{ c | c | c | c | c |}
&   Q_{Sh} & Q_{Sl} & Q_{Ih} & Q_{Il} \\\hline
Q_{Sh} & H\cdot(m_{{Q_{Sh}}}+m_{{Q_{Sl}}}) & L \cdot (m_{{Q_{Sh}}}+m_{{Q_{Sl}}}) & H \cdot (m_{{Q_{Ih}}}+m_{{Q_{Il}}}) & L \cdot (m_{{Q_{Ih}}}+m_{{Q_{Il}}}) \\\hline
Q_{Sl} &  H\cdot(m_{{Q_{Sh}}}+m_{{Q_{Sl}}}) & L \cdot (m_{{Q_{Sh}}}+m_{{Q_{Sl}}}) & H \cdot (m_{{Q_{Ih}}}+m_{{Q_{Il}}}) & L \cdot (m_{{Q_{Ih}}}+m_{{Q_{Il}}})  \\\hline
Q_{Ih} & H \cdot ir & L\cdot ir & H\cdot ii & L\cdot ii \\\hline
Q_{Il} & H\cdot ir & L\cdot ir & H\cdot ii & L\cdot ii  \\\hline
\end{array}
$$
\caption{IDTMC transition probability matrix function $\widehat{\otm}(\vct{m})$, for $\vct{m}$ in $\us{4}$.\label{ex:RPTM}}
\end{figure}

\section{Conclusions}
\label{Concl}
\Abf{}~\cite{CLM16a} is a language for a predicate-based front-end of \FlyFast{,}
an on-the-fly mean-field model-checking tool. In this paper we presented a simplified  version of
the translation proposed in~\cite{CLM16a} together with an approach for
model reduction that can be applied to the result of the translation.
The approach is
based on probabilistic bisimilarity for inhomogeneous DTMCs.
An example of application of the  procedure has been shown. The implementation of
a compiler for \Abf{}  mapping the language to \FlyFast{} is under development as an add-on
of \FlyFast{.}
We plan to apply the resulting extended tool to more as well as more complex models, in order to
get concrete insights on the practical applicability of the framework and on the actual limitations
imposed by the restrictions necessary for exploiting bisimilarity-based state-space reduction.
Investigating possible ways of relaxing some of such restrictions will also be an interesting
line of research.\\$ $\\
\emph{Acknowledgments}
Research partially funded by EU Project
n.~600708 \emph{A Quantitative Approach to Management and Design of
Collective and Adaptive Behaviours} (QUANTICOL).

\bibliographystyle{eptcs}
\bibliography{LatMasQAPL17USB}

\newpage
\appendix
\section{Appendix}
\label{Dec1}

\noindent
{\bf Proof of Proposition~\ref{prop1}}
$ $\\
$\Leftarrow$: Trivial.\\
$\Rightarrow$:\\
We first prove that $a=b$:

\noindent
$
\deriv
\forall m_1,\ldots,m_S. A(m_1,\ldots,m_S) = B(m_1,\ldots,m_S)
\hint{\Rightarrow}{Logic}
A(0,\ldots,0) = B(0,\ldots,0)
\hint{\Rightarrow}{Def. of  $A(m_1,\ldots,m_S)$ and $B(m_1,\ldots,m_S)$}
a = b
$

\noindent
Now we prove  that  $a_{ii} = b_{ii}$ for $i=1,\ldots,S$:\\

\noindent
$
\deriv
\forall m_1,\ldots,m_S. A(m_1,\ldots,m_S) = B(m_1,\ldots,m_S)
\hint{\Rightarrow}{Take the $S-tuple$  $(\bar{m}_1,\ldots,\bar{m}_S)$ where $\bar{m}_k= 1$ if $k=i$ and $0$ otherwise}
A(\bar{m}_1,\ldots,\bar{m}_S) = B(\bar{m}_1,\ldots,\bar{m}_S)
\hint{\Rightarrow}{Def. of  $A(m_1,\ldots,m_S)$ and $B(m_1,\ldots,m_S)$}
a_{ii}+a = b_{ii}+b
\hint{\Rightarrow}{$a=b$ (see above)}
a_{ii}= b_{ii}
$

\noindent
Finally we prove that $a_{ij} = b_{ij}$, for $i,j=1,\ldots,S, j>i$:\\

\noindent
$
\deriv
\forall m_1,\ldots,m_S. A(m_1,\ldots,m_S) = B(m_1,\ldots,m_S)
\hint{\Rightarrow}{$(\tilde{m}_1,\ldots,\tilde{m}_S)$ where $\tilde{m}_k = 0.5$ if $k \in \SET{i,j}$ and $0$ otherwise}
A(\tilde{m}_1,\ldots,\tilde{m}_S) = B(\tilde{m}_1,\ldots,\tilde{m}_S)
\hint{\Rightarrow}{Def. of  $A(m_1,\ldots,m_S)$ and $B(m_1,\ldots,m_S)$}
0.25 a_{ii} + 0.25 a_{ij} + 0.25 a_{jj} + a = 0.25 b_{ii} + 0.25 b_{ij} + 0.25 b_{jj} + b 
\hint{\Rightarrow}{$a=b$ (see above)}
0.25 a_{ii} + 0.25 a_{ij} + 0.25 a_{jj} = 0.25 b_{ii} + 0.25 b_{ij} + 0.25 b_{jj}
\hint{\Rightarrow}{Algebra}
a_{ii} +  a_{ij} + a_{jj} =  b_{ii} + b_{ij} + b_{jj}
\hint{\Rightarrow}{$a_{ii} = b_{ii}$ and $a_{jj}= b_{jj}$ (see above)}
a_{ij}   =  b_{ij} 
$
$ $\hfill$\bullet$

\end{document}